\documentclass[10pt, twoside,leqno,twocolumn]{article}
\usepackage{ltexpprt}
\usepackage[utf8]{inputenc}
\usepackage{graphicx}
\usepackage{hyperref}
\usepackage{bbm}
\usepackage{algorithm}
\usepackage[margin=1in]{geometry}
\usepackage{amsmath}
\usepackage{epsfig, color, graphicx}
\usepackage{enumerate}
\usepackage{hyperref}

\begin{document}

\title{\Large Detecting stock market colluding groups with spectral clustering }
\author{Suneel Sarswat\thanks{Tata Institute of Fundamental Research.} \\
\and
Kandathil Mathew Abraham\thanks{Government of Kerala.}
\and
Subir Kumar Ghosh\thanks{Ramakrishna Mission Vivekananda University.}}
\date{}

\maketitle

\begin{abstract} \baselineskip=10pt We provide the first application of spectral clustering to detecting colluding groups in a stock market. For ensuring market efficiency, stock market regulators take strict action against colluding groups who manipulate the price of stocks and create artificial liquidity. In this work, we show how to use existing machine learning techniques for algorithmically detecting suspected colluding groups based on the stock market data. A key contribution of this work is in appropriately defining `closeness' between two traders that takes into account various factors like common traders and frequency, volume and price of a stock traded between them. In an earlier work, Apte and Palshikar (2008) have applied non-spectral clustering technique to the problem where they use the volume of the trades as a measure of closeness. Our experience with real data suggests that including other factors in measuring closeness helps detect colluding groups that will otherwise go undetected. We 
demonstrate the effectiveness of our algorithm by detecting clusters in a random Erd\H{o}s-R\'enyi graph with hidden clusters.\end{abstract}


\section{Introduction}

\subsection{Trading in a stock exchange}
A stock exchange provides a platform for people to trade stocks of companies that are listed in the exchange. Suppose a potential buyer $X$ intents to buy a stock $S$. So $X$ offers or bids a price for one unit of stock of $S$. A potential seller $Y$, who intents to sell $S$, offers or asks a price for one unit of stock of $S$. Such offers by potential buyers or sellers are called \emph{orders}. If the bidding price is greater than or equal to the asking price, then a trade takes place.  This means that $Y$ transfers a certain units of $S$ (say, $z$) to $X$, and $X$ pays the total money for the $z$ units to $Y$ as per the matched price. The quantity $z$ is called the volume of the trade. \\

The stock exchange does not reveal the identity of a buyer or a seller. Normally, there are many buyers and sellers for a stock. For any stock $S$, such numbers vary throughout the day. In most markets the incoming buy or sell order is either matched with the existing order or placed in a priority queue, where priorities are based on the price.  For \emph{illiquid} stocks, it is possible that a buyer a one seller plan together and place orders on a predetermined price and quantity so that bidding and asking prices match exactly. Such trades are called synchronized trades. \\

In 2011, Securities Exchange Board of India (SEBI), the stock market regulatory of India, initiated regulatory actions against certain individuals \cite{pabari}. As per the  report \cite{pabari}, these individuals were suspected to be involved in creating substantial volumes, which appear to be artificial in nature, executing synchronized and structured trades. This group of individuals was also found to be increasing or maintaining prices and providing misleading signals to the market by artificially injecting volumes in certain stocks and also contributing to the price movement. Further, SEBI observed that such trades appear to be taking place in an unbridled manner. These traders also trade with other non-colluding persons. \\
  
In this paper, we present an algorithm for this problem of identifying such groups of individuals in an efficient manner. Throughout the paper, such groups are referred to as collusion/colluding groups.

\subsection{Problem formulation}
Trading in a stock exchange can be represented as a simple undirected weighted graph $G=(V,E)$, where each vertex of $V$ represents a trader in a stock exchange, and there is an edge between two vertices $v_i$ and $v_j$ of $V$ if (i) there is a trade between $v_i$ and $v_j$ or (ii) there is a trader who has traded with both $v_i$ and $v_j$. Every edge $(v_i, v_j)$ of $G$ is assigned a weight $w_{ij}$. The parameters such as price movement, number of trades, total volume, commonality between every pair $(i,j)$ are used to compute the weights. Since a collusion group is expected to be closely connected through trades, the corresponding subsets of vertices of $G$ are called \emph{clusters} in $G$. The problem of identifying collusion groups in a stock exchange reduces to the problem of identifying clusters in $G$.

\subsection{Our approach}
For detecting collusion groups, graph clustering methods have been used earlier by Palshikar and Apte \cite{apte}, and Islam et. al \cite{islam}. Their algorithms use total volume to compute the weights between two traders, and these algorithms have been tested on small simulated data. For the first time, we apply spectral clustering technique  to the problem. Spectral clustering has been successfully used to find communities in graphs by White and Smyth \cite{white2005spectral}. Moreover, for defining closeness between two vertices of the graph, we use a function to assign weights on edges, where the function is defined in terms of volumes, number of transactions between two individuals, price movements, and commonality between traders. Furthermore, we have used objective metric $Q$ for choosing the number of colluding groups proposed by Newman and Girvan \cite{newman2004finding}. Our algorithm is easy to implement, and it is tested on actual data of SEBI, showing a good performance in practice. Note 
that our graph is very large compared to the graphs used in the earlier works. \\

In the next section, we present a spectral clustering technique used in this paper for locating collusion groups. In Section $3$, we experiment on the data and present the results.

\section{Spectral Clustering}

Spectral clustering is one of the well known modern clustering techniques, used for separating out big data in groups based on closeness. Let $W$ represent a weighted adjacency matrix of a weighted graph $G=(V,E)$ as defined earlier. Let $A$ and $B$ be two disjoint subsets of $V$. Let $W(A)$ and $W(B)$ denote the sum of weights of edges of graph induced by $A$ and $B$ respectively. Let $W(A,B)$ denote the sum of weights of edges between $A$ and $B$. It is easy to see that if $A$ and $B$ are the two different collusion groups of $G$, then $W(A,B)$ should have very low value, whereas both $W(A)$ and $W(B)$ should have high values. Intuitively, $W(A)$ measures closeness amongst the vertices of $A$. So, it is natural to look for subsets $A$ and $B$ such that $W(A)$ and $W(B)$ are maximized and $W(A,B)$ is minimized. Formally, we wish to locate two such subsets $A$ and $B$, such that ${W(A,B) \over W(A)}$ and ${W(A,B) \over W(B)}$ together have low values. \\

Suppose, we add ${W(A,B) \over W(A)}$ and ${W(A,B) \over W(B)}$ instead of considering them separately. So, we get a equation called \emph{MinMaxCut}, which was introduced by Ding et. al \cite{minmax}.

\begin{equation}
\label{eq3}
  MinMaxCut(A,B)=\Bigl({W(A,B) \over W(A)}+{W(A,B) \over W(B)}\Bigl)
\end{equation}

The choice of $A$ and $B$ for which $MinMaxCut(A,B)$ achieves its minimum can be considered as two clusters. This is one method of identifying clusters which we use in this paper. There are other methods for computing clusters such as \emph{RatioCut} \cite{ratio}, \emph{NormalizedCut} \cite{norm} etc.  For our problem of identifying collusion groups, traders in the same cluster have more transaction with each other unlike the traders between the different clusters. The Eq.(\ref{eq3}) captures these properties than other methods. The Eq.(\ref{eq3}) can be generalized to Eq.(\ref{eq4}) for $k$ clusters as follows \cite{tutorial}.

\begin{align}
\label{eq4}
  & MinMaxCut(A_1,A_2, \ldots, A_k) = \nonumber \\
  &\hspace{1cm} \sum_{t=1}^{k} \Bigl({W(A_t,\bar{A_t}) \over W(A_t)}\Bigl)
\end{align}

The choice of $A_1,A_2, \ldots, A_k$, which minimizes Eq.(\ref{eq4}), gives $k$ different clusters. However, the problem of finding such $k$ subsets of vertices is NP-hard  \cite{hard}. Ding et. al \cite{minmax} showed that the problem in Eq. \ref{eq4} can be formulated as trace minimization problem with relaxation on the constrains as follows.

\begin{equation}
\label{eq1}
  \min_{{H \in \mathbbm{R}^{n\times k}}}
    Trace(H^{'}LH) \text{ subject to } H^{'}DH=I 
\end{equation}

Here $D$ is a diagonal matrix such that $d_{ii}$ is sum of the weights of edges of vertex $v_i$ and $L=I-D^{-{1 \over 2}}WD^{-{1 \over 2}}$. The solution of the above problem can be obtained as a solution of generalized eigenvalue problem. In this case, the solution $H$ is to find the first $k$ eigenvectors of $L$ i.e. eigenvector associated with the first $k$ smallest eigenvalues of $L$ as the columns of $H$. For converting a real value solution, $k-$means algorithm can be used on the rows of $H$ to obtain discrete $k$ clusters \cite{tutorial}. 

\subsection{Number of clusters}

To find the number of clusters in $G$ we use the modularity function $Q$ proposed by Newman and Girvan \cite{newman2004finding}. It is defined as follows.
\begin{align}
\label{q}
 Q_k=\sum_{c=1}^{k}{\Bigl[{{\sum_{i \in A_c, j \in A_c}{w_{ij}}} \over {\sum_{i \in V, j \in V}{w_{ij}}}} - \bigl({{\sum_{i \in V, j \in A_c}{w_{ij}}} \over {\sum_{i \in V, j \in V}{w_{ij}}}}\bigr)^2\Bigr]}
\end{align}

The optimal number of clusters $k$ can be achieved by finding the value of $k$ for which $Q_k$ is maximized. 

\section{Our Algorithm}

\subsection{Computing edge weights}

Let us first understand the common features of colluding groups. Assume that two traders $X$ and $Y$ belong to a collusion group. It has been observed that  $X$ and $Y$ generally trades several times between them on the same stock $S$ within a reasonable period of time $d$. Furthermore, during these trades, they tend to trade a  large quantity of $S$. Sometimes, they even trade at a very high or low price from the last trades, if the purpose is to manipulate the price of the stock. In addition, $X$ and $Y$ may even trade through a set of intermediate traders. \\

Let $T_{ij}$ be the total number of trades of $S$ between traders corresponding to $v_i$ and $v_j$ during  $d$. Observe that $T_{ij}$ can be zero if the corresponding traders have not traded $S$ during $d$. Let $T_{max}$ (or $T_{min}$) denote the maximum (respectively, minimum) value of $T_{ij}$ for all pair $(i,j)$. So, $T_{min}\le T_{ij} \le T_{max}$. Since $T_{ij}$ is expected to be close to $T_{max}$ for two traders in a collusion group, the value of the ratio $ {T_{i,j}-T_{min}} \over {T_{max}-T_{min}}$ can be used to assess their closeness. Analogously,  $ {V_{i,j}-V_{min}} \over {V_{max}-V_{min}}$ and $ {P_{i,j}-P_{min}} \over {P_{max}-P_{min}}$ are computed to assess their closeness using volumes and prices respectively. Note that there can be multiple prices for the multiple trades between the two traders. So, $P_{ij}$ is volume weighted average price between $v_i$ and $v_j$. Let $N_i$ and $N_j$ be the set of neighbors of $v_i$ and $v_j$ respectively. To incorporate the intermediate traders in the 
$w_{i,i}$, the common neighbors ${N_i}\cap {N_j}$ are expected to be very close to the total number of neighbors ${N_i}\cup {N_j}$. Note that $N_i$ contains $v_i$ and $N_j$ contains $v_j$. Hence, we have the following formula for computing $w_{i,j}$.
\begin{align}   
\label{w}
w_{i,j} &= \dfrac{1}{4} \left(\dfrac{T_{i,j}-T_{min}}{T_{max}-T_{min}} + \dfrac{V_{i,j}-V_{min}}{V_{max}-V_{min}} \right.\nonumber \\
        &\hspace{1cm} \left. + \dfrac{P_{i,j}-P_{min}}{P_{max}-P_{min}} + \dfrac{|{N_i}\cap {N_j}|}{|{N_i}\cup {N_j}|}\right). 
\end{align}

Observe that the value ${N_i \cap N_j} \over {N_i \cup N_j}$ in the above equation is one if $v_i$ and $v_j$ share all their neighbors. Even if traders corresponds to $v_i$ and $v_j$ are not trading directly but they trade through the same set of traders, the edge between them receives non-zero weights. 

\subsection{Computing Clusters}
Now, we present the main steps of the algorithm for locating $k$ clusters in a graph $G$. 

\begin{algorithm}[H]
\caption{Spectral Graph Clustering}
\label{alg1}
Input: $G$ and $k$. \\
For $2 \le k \le n$. 
Construct $W$ from $G$. \\
Compute $D$ and $L$.\\
Compute the first $k$ eigenvectors of $L$ and construct matrix $Q \in \mathbbm{R}^{n\times k}$ by placing $k$ Eigenvectors as columns of $Q$.\\
Construct matrix $U$ from $Q$ by assigning $u_{i,j}={{q_{i,j}}\over{\sqrt{\sum_j{q_{i,j}^2}}}}$.\\
For $i = 1, \ldots , n$, let $y_i \in \mathbbm{R}^k$ be the vector corresponding to the $i^{th}$ row of $U$. \\
Cluster the points $(y_i)_{i=1,\ldots,n}$  with the $k$-means algorithm into clusters $C_1, \ldots , C_k$.\\
Compute $Q_k$. \\
Pick the corresponding partition which maximize $Q_k$. \\
Output: Clusters $A_1, A_2, \ldots, A_k$ with $A_i \in \{v_j \mid y_j\in C_i\}$.

\end{algorithm}

\section{Experiment and Results}

\subsection{Market data}

 Market data consist of all trades of every stock for the entire period in the two main exchanges in India, namely, National Stock Exchange and Bombay Stock Exchange. The total number of trades for a period of one year are more than a billion for all the stocks. Each trade data contains all information or parameters consisting of (i) codes of the two traders, (ii) date and time of the trade, (iii) stock name (iv) traded price of the stock, and (v) traded volume of the stock. \\
 
 Consider a situation where two traders $X$ and $Y$ have same address or same telephone number or \emph{off-market} transactions or any other common parameter. Off-market trades are those trades where stocks are transfered from one account to another account directly without exchange. These trades indicates that two individuals know each other and are trading knowingly. These informations can be used by the regulators to verify the validity of colluding group. \\

 Using Algorithm \ref{alg1}, we analyzed trade data for the period of 17 months. We observed that some stocks had colluding groups. In many such stocks, there is usually only one colluding group per stock, i.e., $k=1$. After Algorithm \ref{alg1} identified a cluster $C$ for a company, the regulators of the stock markets verified whether $C$ was indeed a colluding group, using the parameters of traders in $C$ mentioned above. The verification showed that $C$ included most members of the colluding groups. Since the details of the results are classified, here we use publicly available data and simulated data to demonstrate the performance of our algorithm. For the simulated data we assumed that weights follows uniform distribution and the parameters of actual data is considered for the simulated data.

\subsection{Simulated data}

We construct a random graph $G(V,E)$ which is used as an input to Algorithm \ref{alg1}. Let $G(n,p)$ be a random graph of size $n$ such that there is an edge between any two vertices of the graph with probability $p$ \cite{erdds1959random}. Initialize $G(V,E)$ by $G(n,p)$. Choose any two subsets $C_1 \subset V$ and $C_2 \subset V$ of size $n_1$ and $n_2$ respectively. Add edges in $C_1$ and similarly in $C_2$ such that graph induced by $C_1$ is  $G(n_1,p_1)$ and $C_2$ is  $G(n_2,p_2)$ with $p_1, p_2 > p$. The weight matrix $W$ of $G$ is constructed such that $w_{i,j}$ follows uniform distribution i.e. $w_{i,j} \sim U(0,1)$ if $i,j \in C_1$ or $i,j \in C_2$ else $w_{ij} \sim U(0,b)$ where $b \le {1 \over 2}$. Then $G$ is used as an input to the Algorithm \ref{alg1}. The experiments are repeated for many times for various values of $n, n_1,n_2, p, p_1, p_2, b$. The clusters $A$, $B$ and $C$ identified by the algorithm are compared with $C_1$ and $C_2$, and the results are shown  below. \\

\begin{figure}[H] 
 \begin{center}
 \centerline{\hbox{\psfig{figure=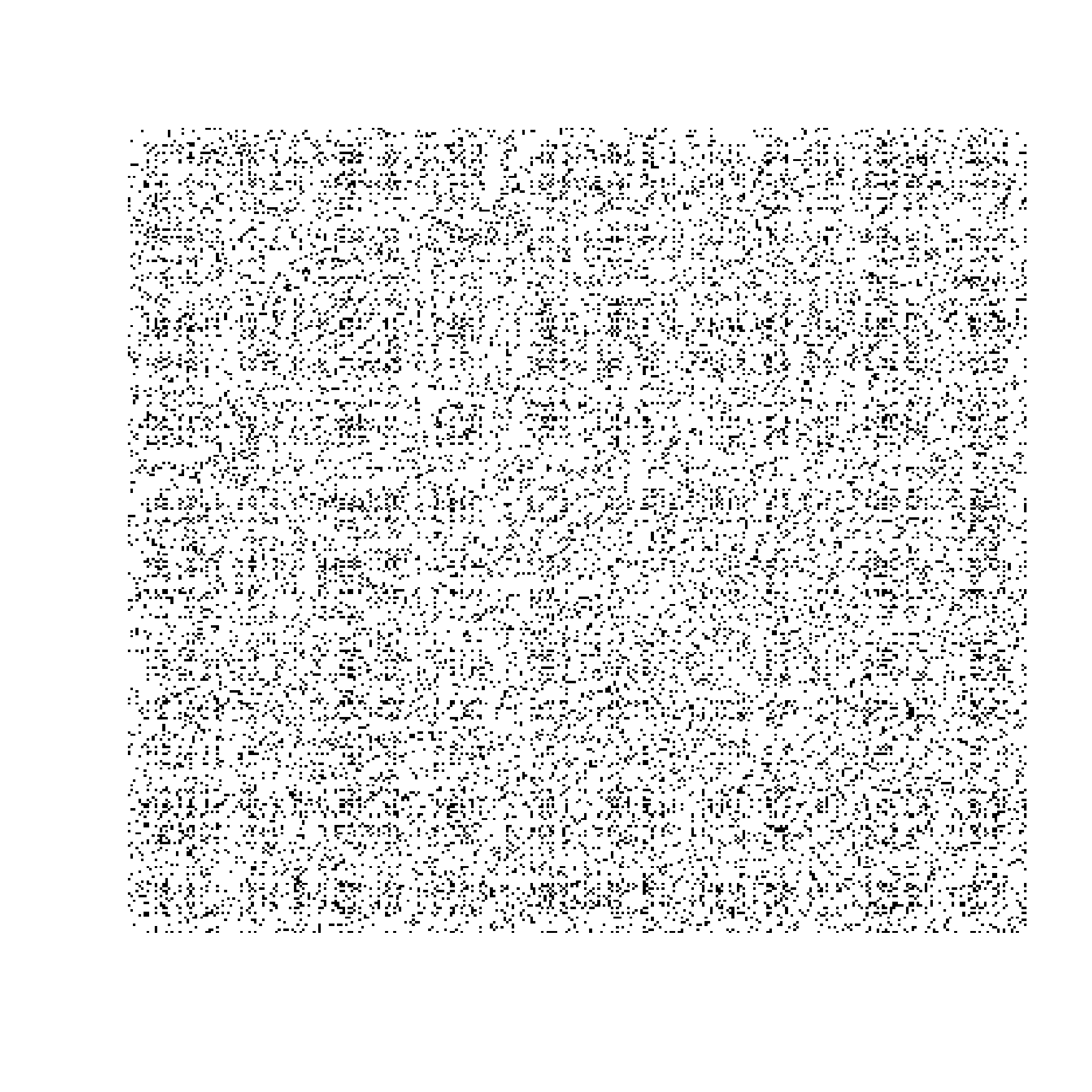,width=.6\hsize}}}
 \caption{Adjacency matrix of $G(V,E)$ for $n=335$.}
 \label{adja_b}
 \end{center}
 \end{figure}

\begin{figure}[H] 
 \begin{center}
 \centerline{\hbox{\psfig{figure=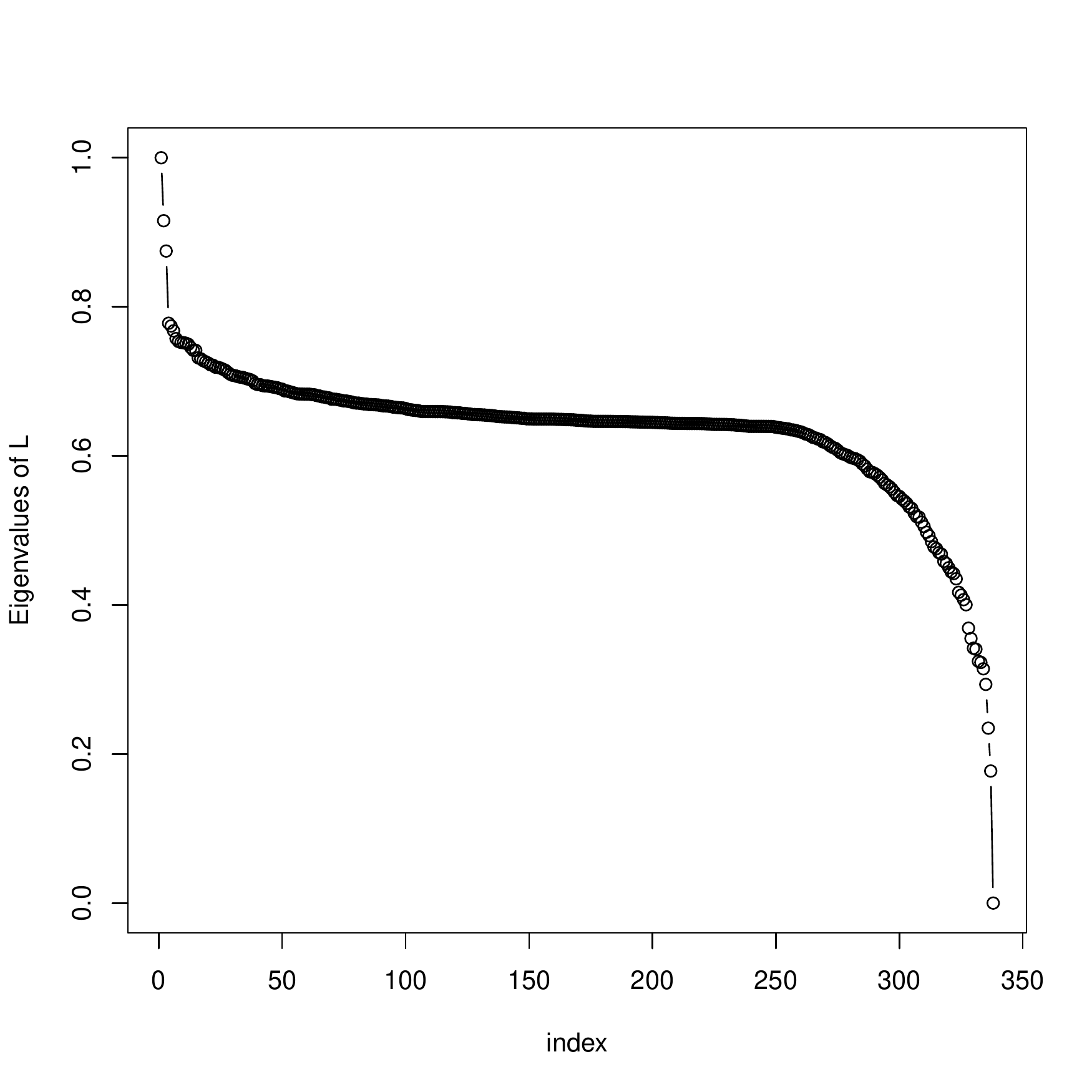,width=1\hsize}}}
 \caption{The eigenvalues of $L$ for $n=335$. The three isolated dots on left indicates three clusters in the graph \cite{tutorial}.}
 \label{eigen_values}
 \end{center}
 \end{figure}

Figure \ref{adja_a} is a pictorial image of adjacency matrix of $G(V,E)$ with $n=335$, $p=.1$,$p_1=.7$,$p_2=.7$,$n_1=50$, $n_2=60$ and $b=.4$. The ordering of second eigenvector of $L$ is used to the pictorial image  of reordered adjacency matrix is presented in Figure \ref{adja_b}. Figure \ref{eigen_values} is the plot of eigenvalues of $L$.

 \begin{figure}[H] 
 \begin{center}
 \centerline{\hbox{\psfig{figure=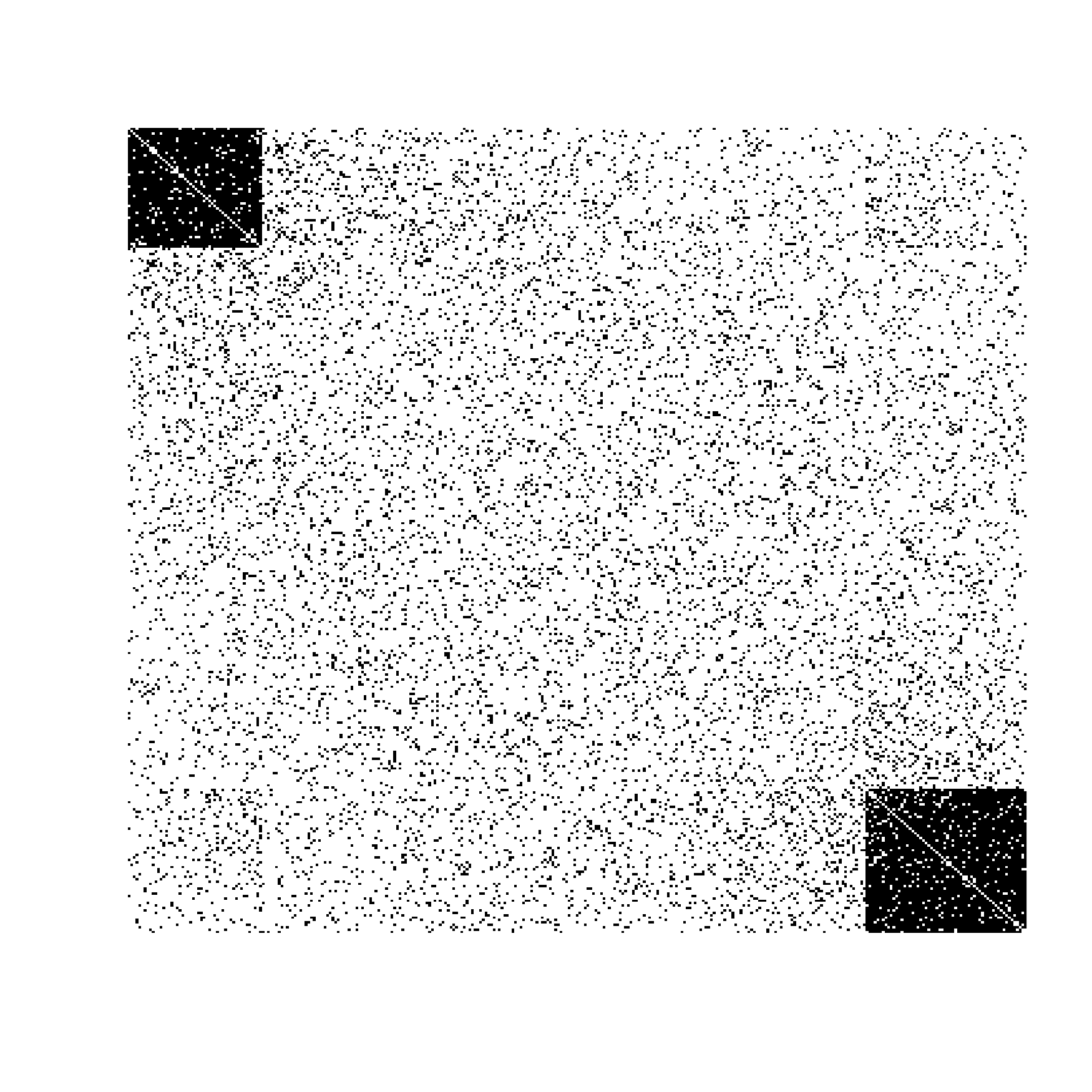,width=.7\hsize}}}
 \caption{The adjacency matrix is obtained after running the algorithm and shifting rows and columns using orders of second eigenvector of $L$. In this matrix two colluding groups are clearly visible.}
 \label{adja_a}
 \end{center}
 \end{figure}

\subsection{Publicly available data}

\begin{figure}[H] 
 \begin{center}
 \centerline{\hbox{\psfig{figure=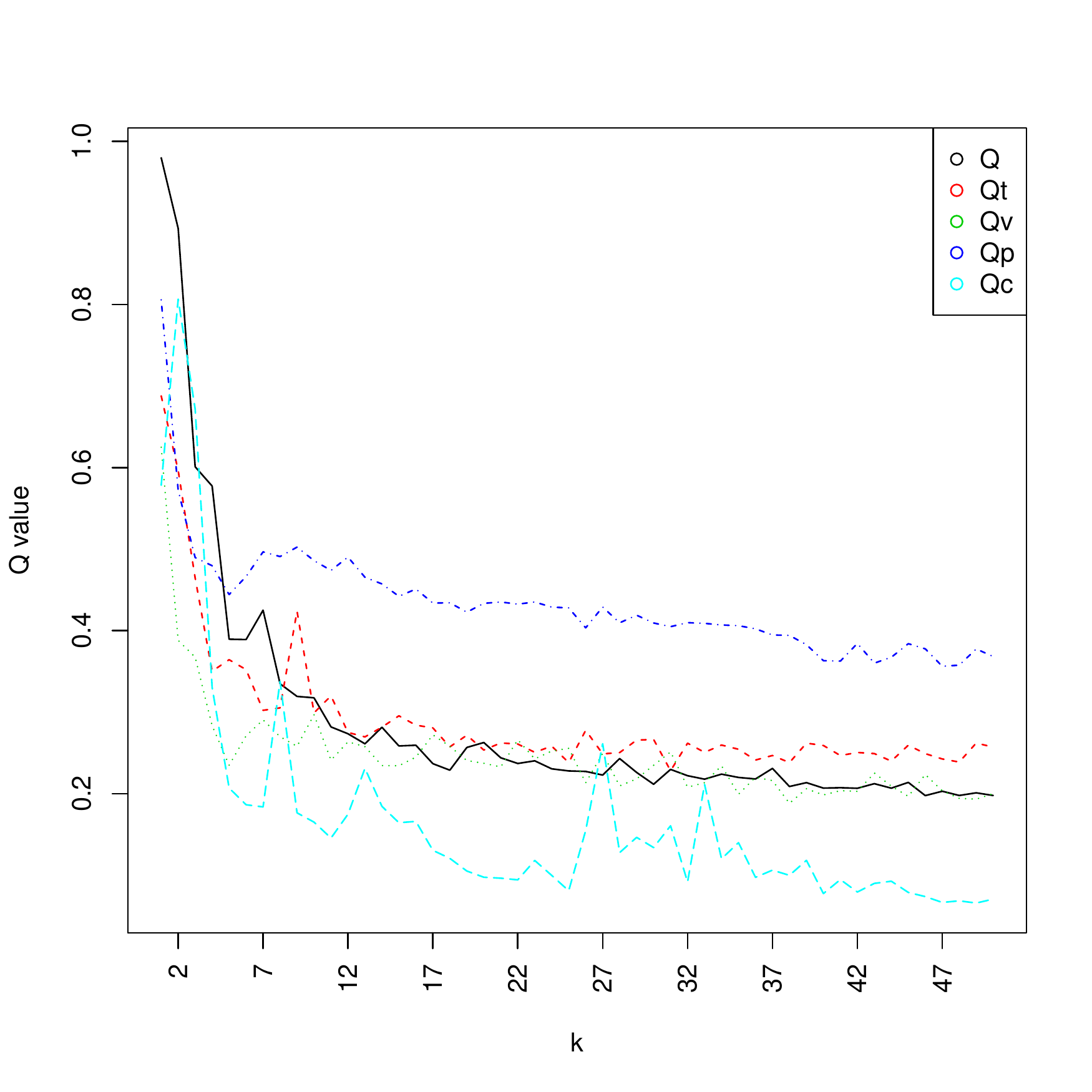,width=1\hsize}}}
 \caption{In this figure the $Q$ value is computed using our algorithm. The value $Qt$ corresponds to the $Q$ value in equation \ref{q} when the number of transactions is the only parameter to compute weights in the graph i.e. only first term in equation \ref{w} is considered to compute the weights. Similarity $Qv$, $Qp$ and $Qc$ corresponds to volume, price and commonality i.e. second, third and forth terms in equation \ref{w} respectively. The total number of clusters in this example is $2$.}
 \label{Q_values}
 \end{center}
 \end{figure}

We have used the stock trade data of Bombay Stock Exchange for the period of 2011. These data contains information such as traders order ids, volume, price and time of the trade. The order id is a $17$ digit number and we assume that  the first $6$ digits corresponds to the id of an individual for our experiments. In Figure \ref{Q_values} we show the $Q$ values for various values of $k$ for a stock. We have also compared the results obtained by this algorithm in case only one of the parameter is used i.e. number of transactions, volume, price or commonality.

\subsection{Financial Implications of price manipulation}
Price manipulation in stock market may impact the other financial institutions. For example, many banks provide loan against stocks and the amount of loan depends on the current price of the stock. If the price of a stock is manipulated to make it higher, then the sanctioned loan amount from a bank can be increased. In case of the default of the loan, the loan becomes non performing asset to the bank since it is difficult for the bank to sell the stock and recover full amount of the loan. \\

There is another reason which may motivate traders to manipulate the price of a stock. In many economy, a short-term capital losses can be set off against long/short term capital gains to compute the taxable income. To understand this, let us consider two investors $A$ and $B$. Assume that $A$ has some long term taxable income (say, $y$) from some business and $B$ has incurred short term loss of $y$. Suppose $A$ buys some stock $S$ from $B$ at a very highly manipulated price. After the manipulation is stopped, the price of $S$ reduces and $S$ is sold back to $B$ by $A$ at lower price. This brings a short term loss, say $x$, in the account of $A$ and a short term gain of $x$ in the account of $B$. This way $A$ can save his taxes through losses of $B$ by $y-x$ and $B$ still does not have to pay any tax. So the tax, which otherwise could have gone to the government, gets converted into black money.

\section{Conclusion}

Detecting colluding group is a challenge for the regulators of the securities markets. So, an automated surveillance system which detects the suspect group of traders involved in colluding is an important problem. In this work we have presented an algorithm which detects such groups. Simulated data is constructed here in such a way that it resembles the actual data. Naturally, our Algorithm \ref{alg1} also perform well on the simulated data. Hence, our algorithm is very practical for identifying collusion groups.\\

Spectral clustering can be used for other finance problems as well. For example, finding the clusters of the stocks which are similar. This can be used to diversify a portfolio. It can also be used to classify the mutual funds into various categories. One technique to formulate the problem is to use weights between two mutual funds as Jaccard similarity coefficient, where a mutual fund can be considered a set of stocks. \\
 
\noindent {{\Large\bf Acknowledgements}} \\

The authors thank National Institute of Securities Markets for providing trading data for the analysis. The author gratefully acknowledges helpful comments and suggestions of Lata Chari, Mohit Garg, Daya Gaur and Bodhayan Roy.

\end{document}